\documentclass[aps,twocolumn,showpacs]{revtex4}
\usepackage{graphicx}
\usepackage{bm}
\usepackage{amsmath}
\begin{document}

\title{States of Local Moment Induced by Nonmagnetic Impurities in Cuprate
Superconductors}

\author{Yan Chen and C. S. Ting}

\affiliation{Texas Center for Superconductivity and Department
of Physics, University of Houston, Houston, TX 77204}

\begin{abstract}

By using a model Hamiltonian with $d$-wave superconductivity and
competing antiferromagnetic (AF) orders, the local staggered
magnetization distribution due to nonmagnetic impurities in
cuprate superconductors is investigated. From this, the net
magnetic moment induced by a single or double impurities can be
obtained. We show that the net moment induced by a single impurity
corresponds to a local spin with $S_z$=0, or 1/2 depending on the
strength of the AF interaction and the impurity scattering. When
two impurities are placed at the nearest neighboring sites, the
net moment is always zero. For two unitary impurities at the next
nearest neighboring sites, and at sites separated by a Cu-ion
site, the induced net moment has $S_z=$0, or 1/2, or 1. The
consequence of these results on experiments will be discussed.

\end{abstract}

\pacs{75.20.Hr, 74.25.Jb, 72.10.Fk, 71.55.-i}

\maketitle

The nonmagnetic impurity effect in high temperature
superconductors (HTS) has attracted significant interest both
experimentally and theoretically for many years. The induction of
local magnetic moment would be expected due to the competition
between spin magnetism and superconductivity in these systems.
Nuclear magnetic resonance (NMR) measurements in
YBa$_2$Cu$_3$O$_{7-x}$ have indicated that nonmagnetic Zn/Li
impurities, enhance antiferromagnetic correlation and staggered
magnetic moment is induced on the Cu ions in the vicinity of the
impurity
sites~\cite{Alloul,mahajan,bobroff99,kilian,mac00,julien00,bobroff01}.
Low-temperature scanning tunneling microscopy (STM) experiments
~\cite{Pan} have directly observed a sharp near zero bias
resonance peak around the Zn impurity atoms on the surface of
superconducting Bi$_2$Sr$_2$CaCu$_2$O$_{8+x}$ (BSCCO). Both local
potential scattering~\cite{balatsky,Zhu01,flatte,Tsuchiura,Zhu02}
and Kondo impurity scattering~\cite{Zhu03,sachdev,Dai1} in
$d$-wave superconductors has been theoretically investigated.
Recently Wang and Lee~\cite{WangLee} studied the formation of the
local moment near a nonmagnetic impurity in HTS using  a
renormalized mean field theory of the $t-J$ model. They tried to
reconcile the formation of local moment around a unitary
nonmagnetic impurity and strong zero bias resonance in the local
density of states (LDOS) spectrum.

On the other hand, quantum interference effect among multiple
impurities in HTS draws considerable attention in recent
years~\cite{Flatte,Morr1,Morr2,LZhu,Andersen,Atkinson}. The energy
dependent of the spatial distribution of LDOS spectrum changes
remarkably by varying the distance and orientations among the
impurities.  To our knowledge, so far there exists no study
considering the local moment formation due to quantum interference
effect among multiple impurities in HTS. In order to study this
and related problems, we apply an effective model Hamiltonian
defined on a square lattice with a nearest neighboring (n.n.)
interaction $V_{DSC}$ to simulate the $d$-wave superconductivity
(DSC) and an onsite Coulomb interaction $U$ to represent the
competing antiferromagnetic (AF) order. In this paper, the local
magnetic moment distribution induced around a single impurity will
be numerically studied first. We show that the net moment induced
around the impurity can be attributed to a local spin with
$S_z$=0, or 1/2 depending on the impurity strength and the value
of $U$, and we shall also present a phase diagram for the net
moment formation.  With multiple impurities, we expect that
quantum interference effect should exhibit itself. When two
impurities are placed at the n.n. sites, our numerical result
indicates that the net induced moment is always zero, regardless
of the value of $U$, the impurity strength, and doping. For two
unitary impurities at the next n.n. sites and at sites separated
by a Cu-ion site, we demonstrate that the net moment induced by
them can be represented by a local spin with $S_z$=0, or 1/2, or 1
depending on the strength of the AF interaction or doping.

We begin with a phenomenological model Hamiltonian in a
two-dimensional plane, in which both the DSC and the competing AF
or the spin density wave (SDW) order are taken into account:
\begin{eqnarray}
H&=&-\sum_{i,j,\sigma} t_{ij} c_{i\sigma}^{\dagger}c_{j\sigma}
+\sum_{i,\sigma}( U n_{i {\bar {\sigma}}} + \epsilon
\delta_{i_m,i} -\mu) c_{i\sigma}^{\dagger} c_{i\sigma} \nonumber
\\ &&+\sum_{i,j} ( {\Delta_{ij}} c_{i\uparrow}^{\dagger}
c_{j\downarrow}^{\dagger} + h.c.)\;,
\end{eqnarray}
where $\epsilon$ is the on-site impurity strength, $i_m$ is the
impurity site, $\mu$ is the chemical potential. The hopping term
includes the n.n. hopping $t$ and the next n.n. hopping
$t^{\prime}$. The staggered magnetization and DSC order in
cuprates are defined as $M_i^{s} = (-1)^{i} \langle c_{i
\uparrow}^{\dagger} c_{i \uparrow} -c_{i \downarrow}^{\dagger}c_{i
\downarrow} \rangle$ and $\Delta_{ij}=V_{DSC} \langle
c_{i\uparrow}c_{j\downarrow}-c_{i \downarrow} c_{j\uparrow}
\rangle /2$. The mean-field Hamiltonian (1) can be diagonalized by
solving the resulting Bogoliubov-de Gennes equations
self-consistently
\begin{equation}
\sum_{j} \left(\begin{array}{cc} {\cal H}_{ij,\sigma} &
\Delta_{ij} \\ \Delta_{ij}^{*} & -{\cal H}_{ij,\bar{\sigma}}^{*}
\end{array}
\right) \left(\begin{array}{c} u_{j,\sigma}^{n} \\
v_{j,\bar{\sigma}}^{n}
\end{array}
\right) =E_{n} \left(
\begin{array}{c}
u_{i,\sigma}^{n} \\ v_{i,\bar{\sigma}}^{n}
\end{array}
\right)\;,
\end{equation}
where the single particle Hamiltonian ${\cal H}_{ ij,\sigma} =
-t_{ij} +(U n_{i \bar{\sigma}}  + \epsilon \delta_{i_m,i}
-\mu)\delta_{ij}$, and $n_{i \uparrow} = \sum_{n}
|u_{i\uparrow}^{n}|^2 f(E_{n})$, $ n_{i \downarrow} = \sum_{n}
|v_{i\downarrow}^{n}|^2 ( 1- f(E_{n}))$, $ \Delta_{ij} =
\frac{V_{DSC}} {4} \sum_{n} (u_{i\uparrow}^{n}
v_{j\downarrow}^{n*} +v_{i\downarrow}^{*} u_{j\uparrow}^{n}) \tanh
\left( \frac{E_{n}} {2k_{B}T} \right)$, with $f(E)$ as the Fermi
distribution function. The DSC order parameter at the $i$th site
is $\Delta^{D}_{i}= (\Delta_{i+e_x,i} + \Delta_{ i-e_x,i} -
\Delta_{ i,i+e_y} -\Delta_{ i,i-e_y})/4$ where ${\bf e}_{x,y}$
denotes the unit vector along $(x,y)$ direction. Various doping
concentrations can be tuned by varying the chemical potential. In
the present calculation, we set the lattice constant $a$ and
hopping integral $t$ as units, $t^{\prime} =-0.2$ and
$V_{DSC}=1.0$. Due to the localized nature of the impurity states,
the order parameters around impurity are insensitive to the
boundary conditions for large system sizes. The linear dimension
of the unit cell of the vortex lattice is chosen as $N_x \times
N_y = 32 \times 32$. The averaged electron density is fixed at
$\bar {n}=0.85$. The calculation is performed in very low
temperature regime. The supercell techniques are employed to
calculate the LDOS.  The number of the unit cells is $M_x \times
M_y = 20 \times 20$.

Our numerical results for a single impurity show that the local AF
order may be absent around the impurity site for small $\epsilon$
and is present when impurity strength $\epsilon$ becomes larger.
In Fig. 1, we plot three typical spatial distributions of the
staggered magnetization and their corresponding LDOS spectra
around the impurity site. The net local moment is defined as $S_z
= \sum_i S_i^z$. The impurity is situated at $(16,16)$. Panels (a)
and (b)  corresponds to $U=2.0$ and $\epsilon =3$. It is clear
that no AF order induction is shown for such a weak impurity. In
panel (b), the resonance energy at the impurity site (black line)
is less than zero while the resonance energy at the n.n. impurity
site (red line) is greater than zero. The green line is for the
site far from the impurity site. With the increasing of the
impurity strength, the resonance peak height at the impurity site
becomes weaker and its peak shifts to zero energy while the
resonance peak height at the n.n. impurity site turns stronger and
its peak also moves to zero energy. Panels (c) and (d) are for a
strong or unitary impurity $\epsilon=100$ and $U$=2.0.  We find
the induced staggered magnetization reaches a maximum value of
0.08 at the n.n. site of the impurity in Fig. 1(c), and the local
static AF fluctuation extends over several lattice sites from the
impurity but there is no net induced moment(or $S_z=0$). The
remarkable enhancement of zero bias resonance peak at the n.n.
sites is shown in panel (d), reflecting the characteristics of a
unitary impurity. The presence of weak local AF order results also
a rather weak splitting of resonance peak. In Figs. 1(e) and 1(f)
we show the our results for $U=2.35$ and $\epsilon =100$. A
pronounced two-dimensional SDW order is clearly induced around
this unitary impurity in Fig. 1(e). The staggered magnetization at
the n.n. site reaches to  0.33 and the net induced moment becomes
$S_z=1/2$.
\begin{figure}[t]
\includegraphics[width=8.6cm]{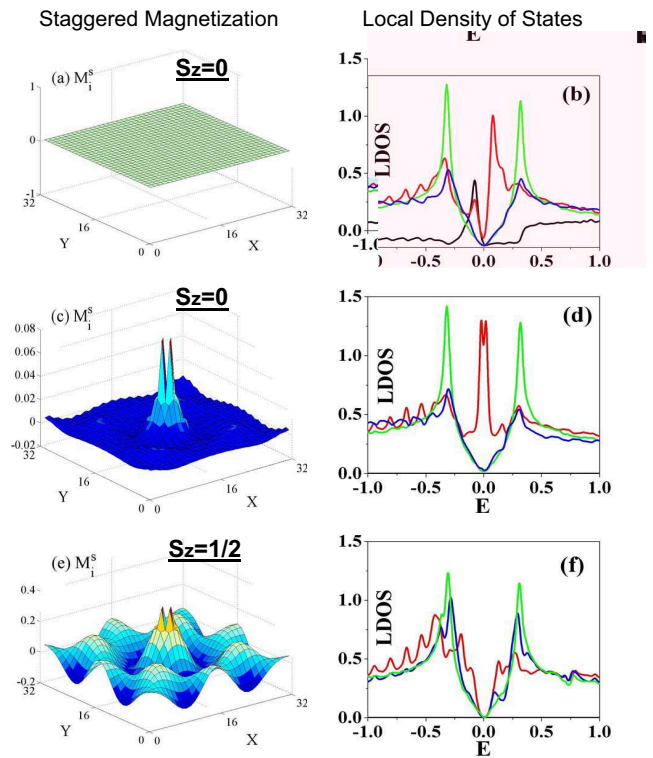}
\vspace{-1.6cm} \caption{\label{Fig1} Spatial variations of the
staggered magnetization $M_{i}^{s}$ [(a), (c) and (e)], and LDOS
spectra [(b), (d) and (f)]. The impurity site is at (16,16). The
black line in panels (b), (d) and (f) is for (16,16), red line for
(17,16), blue line for (17,17) and green line for (16,1). The
upper panels [(a), (b)], the central panels [(c), (d)] and the
lower panels [(e), (f)] are for $(U=2.0, \epsilon=3)$, $(U=2.0,
\epsilon=100)$ and $(U=2.35, \epsilon=100)$, respectively.}
\end{figure}
The zero bias resonance peak of the LDOS at a smaller $U$ with $S_z=0$ shown in
Fig. 1(d) at the n.n. site is substantially  suppressed and become two weak peaks in
the larger $U$ case with $S_z=1/2$(see Fig. 1(e)).

To examine the local moment formation, we present the phase
diagram of $\epsilon$ versus $U$ in Fig. 2.
\begin{figure}[t]
\includegraphics[width=7.5cm]{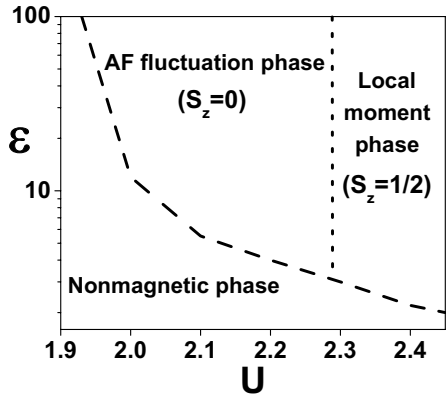}
\vspace{-0.5cm} \caption{\label{Fig2} Phase diagram of $\epsilon$
versus interaction strength $U$ for various  phases near the impurity.}
\end{figure}
It is obvious that the induced net moment ($S_z=1/2$) should  show
up for larger impurity strength $\epsilon$ and stronger AF
interaction $U$, while $S_z=0$ tends to exist for smaller
$\epsilon$ or weaker $U$. In fact, there exist three phase regimes
depending on the magnitude of $U$ value or doping (not shown
here). For small $U$, no AF order (nonmagnetic phase) is induced
around the impurity (see Fig.1(a)). For intermediate $U$, weak
local AF order (AF fluctuation phase) appears (see Fig. 1(c)). In
both cases the net induced magnetic moment corresponds to a local
spin with  $S_z=0$. For larger $U$, a pronounced SDW order is
induced (see Fig. 1(e)) and the net moment becomes $S_z=1/2$
(local moment phase). In the nonmagnetic phase regime and for a
unitary impurity, there is a sharp  zero bias resonance peak in
the LDOS without any splitting~\cite{balatsky}. The splitting is
gradually showing up in the AF fluctuation phase regime. The LDOS
could be substantially suppressed at zero bias and split into two
weak peaks in the third regime with $S_z$=1/2. It is important to
notice that the AF fluctuation phase regime discussed here is
absent in Ref. [17]. The staggered magnetization distribution
around a unitary  impurity like Zn obtained in this paper is in
agreement with the results derived from the NMR experiments
~\cite{Alloul,mahajan,bobroff99,kilian,mac00,julien00,bobroff01}.
The critical value of the $U$ for inducing $S_z=1/2$ is doping
dependent. The underdoped case corresponds to smaller $U_c$. The
$S_z=1/2$ moment due to a unitary impurity is much easier to be
induced in an underdoped sample than in optimally and overdoped
samples, because the amplitude of the induced AF order becomes
increasingly stronger as the hole doping is decreased. Since the
existence of imhomogenity in the HTS sample has been
experimentally confirmed~\cite{Pan2,Lang}, as a result, the number
of induced $S_z=1/2$ moments would be smaller than the number of
unitary impurities like Zn. We also predict that the strong
zero-bias-peak observed by Pan {\em et al.}~\cite{Pan} at the Zn
impurity site should be associated with $S_z=0$ moment in the
overdoped and possibly optimally doped regions. In the underdoped
region, the induced moment would become $S_z=1/2$ where the LDOS
spectrum  should be much suppressed by the induced SDW at  zero
bias.

We next study the quantum interference effect on the local moment
formation due to two unitary impurities. When they are placed at
the n.n. sites (see Fig. 3(a)), our numerical results for the
distribution of the induced magnetization around the impurities
are shown in Figs. 3(b) and 3(c) with respectively $U$=2.2 and
2.4. For $U$=2.0 and 2.35 (not shown here), the induced staggered
magnetizations are also uniformly zero and exactly identical to
that in Fig. 3(b). Although it appears that the staggered
magnetizations due to the two n.n. impurities have exact
cancellation, their influence on the LDOS is still apparent and
the result will not be present here. With $U$=2.4 and no
impurities, one can numerically demonstrate that the staggered
magnetization has a stripe like structure~\cite{ChenTing} with
periodicity 8$a$ which coexists with the DSC.  The presence of the
impurities could pin the stripes but does not modify the overall
stripe-like structure except the magnetizations at or very close
to the impurity sites are altered (see Fig. 1(c)). In all these
cases, the net induced moment has $S_z=0$. This result is very
robust and independent of the value of $U$, impurity strength
$\epsilon$, and doping.
\begin{figure}[t]
\includegraphics[width=8.6cm]{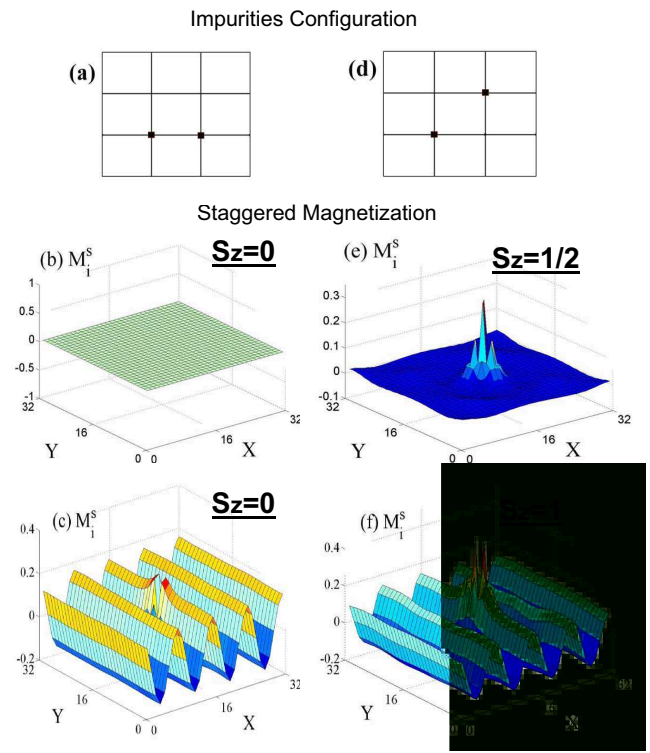}
\vspace{-1.6cm} \caption{\label{Fig3} Impurities configuration
[(a) and (d)] and staggered magnetization $M_{i}^{s}$ [(b), (c)
and (e), (f)]. The left panels [(a), (b), and (c)] and the right
panels [(d), (e), and (f)] are for two n.n. impurities and two
next n.n. impurities, respectively. Panels [(b), (c)] and [(e),
(f)] correspond to $U=2.0$ and $U=2.4$, respectively.}
\end{figure}
For two unitary impurities placed at the next n.n. sites (see Fig.
3(d)), The induced spatial profiles of the staggered magnetization
for $U$=2.0 and 2.4 are respectively shown in Figs. 3(e) and 3(f).
The net moment associated with Fig. 3(e) yields a local spin of
$S_z=1/2$, while that associated with Fig. 3(f) has $S_z=1$. For
$U=2.35$, the induced SDW no longer has the  stripe like structure
and we still obtain $S_z=1$.   The induced net moment has been
shown to have $S_z=0$ when $U$ is less than 1.9.

Finally, we place two unitary impurities at sites separated by a
Cu ion (see Fig. 4(a)). The distributions of the staggered
magnetizations are shown in Figs. 4(b), 4(c) and 4(d) for $U$=1.9,
2.0 and 2.4, and the induced net moments have respectively
$S_z$=0, 1/2 and 1.
\begin{figure}[t]
\includegraphics[width=8.6cm]{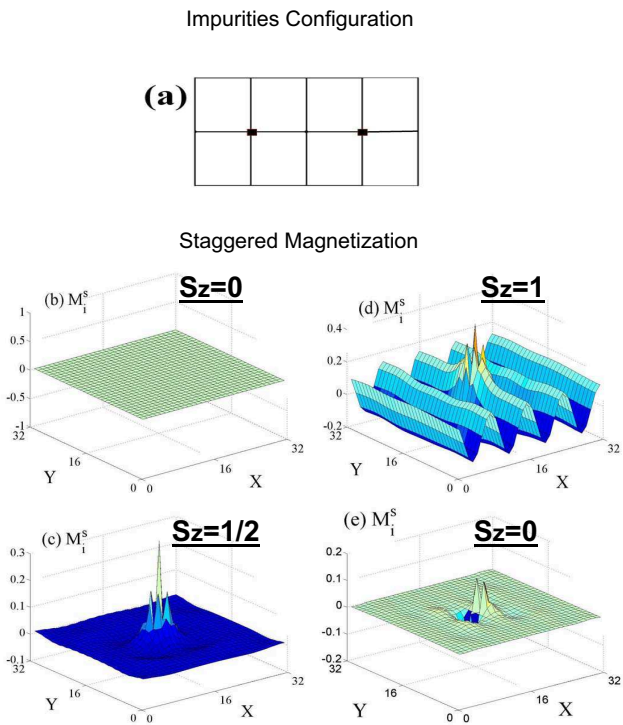}
\vspace{-1.6cm} \caption{\label{Fig4} Unitary impurities
configuration (a) and spatial profiles of staggered magnetization
$M_{i}^{s}$ [(b), (c) ,(d) and (e)] in a $32 \times 32$ lattice.
Their corresponding parameters are (b) ($U=1.9$), (c) ($U=2.0$),
(d) ($U=2.4$) and (e) ($U=2.2, t^{\prime}=-0.3$).}
\end{figure}
In Fig. 4(b), no induced AF order is present for small $U$. The
net moment has $S_z=0$. In Fig. 4(c), a remarkable enhancement of
local AF order is shown at the central Cu site and the DSC order
at this site is suppressed to almost zero. This is a kind of {\em
constructive} interference effect. Its net moment has $S_z=1/2$.
In Fig 4(d), one can clearly observe the pinning of SDW stripes by
impurities. The local magnetization at the central Cu site is also
enhanced. The net moment has $S_z=1$. As shown in Fig. 4(e), if we
choose $U=2.2$ and a different band parameter $t^{\prime}=-0.3$,
the induced local moment associated with two individual impurities
would have opposite polarity and yield a net $S_z=0$. The {\em
destructive} interference effect is shown at the center site where
local AF order is equal to zero. This type of opposite polarities
occurs when two impurities separated further apart even with
$t^{\prime}=-0.2$ as used in the present paper. The net local
moment induced by two impurities spaced by one Cu ion could result
$S_z$= 0, or 1/2 or 1. In other words, quantum phase transitions
may occur among different local moment phases by varying $U$
values or doping. The detailed study of various configurations of
the impurities on the local moment formation and LDOS spectra will
be presented as a future work.

In summary, we have investigated  the induction of the local
moment by a single and double impurities in HTS based on a
phenomenological model with DSC and competing AF orders. By tuning
the impurity potential and the value of $U$, a transition between
various net magnetic moment states may appear. The LDOS has been
calculated for a single impurity. We show that the zero bias
resonant peak obtained next to the unitary impurity site is always
associated with weak $U$ and $S_z=0$. When $S_z=1/2$, the LDOS at
zero bias is suppressed by the gap of the locally induced SDW.
This is consistent with the recent STM experiments~\cite{Lang},
where the zero bias resonant peaks due to Zn impurities are
observed only in the hole rich region, not in the hole poor region
in BSCCO. In addition, the quantum interference effect by two
nonmagnetic impurities has also been studied. Our calculation
predicts the absence of net magnetic moment around two n.n.
impurities, regardless of the values of the impurity strength,
doping, and $U$. This result indicates that the number of induced
$S_z=1/2$ moments is always smaller than the number Zn impurities
even in an underdoped sample where the AF strength is appreciable.
We demonstrate that net magnetic moment around two unitary
impurities placed at the next n.n. sites and sites separated by a
Cu ion could be either 0, 1/2 or 1 depending on the value of $U$.
It is our hope that the present investigation on the local moment
formation may provide useful information for future experimental
test.

We are grateful to Profs. T.K. Lee, S.H. Pan, Z.Q. Wang, G.-q.
Zheng and Dr. J. X. Zhu for useful discussions. This work was
supported by the Robert A. Welch Foundation, by the Texas Center
for Superconductivity at the University of Houston through the
State of Texas.

\end{document}